# Ferroaxial phonons in chiral and polar NiCo$_2$TeO$_6$

V. A. Martinez,[1,*] Y. Gao,[1] J. Yang,[1] F. Lyzwa,[1,2] Z. Liu,[3] C. J. Won,[4] K. Du,[5] V. Kiryukhin,[5] S-W. Cheong,[5] and A. A. Sirenko[1]

[1] *Department of Physics, New Jersey Institute of Technology, Newark, New Jersey 07102, USA*

[2] *NSLS-II, Brookhaven National Laboratory, Upton, New York 11973, USA*

[3] *Department of Physics, University of Illinois at Chicago, Chicago, Illinois 60607-7059, USA*

[4] *Laboratory for Pohang Emergent Materials and Max Plank POSTECH Center for Complex Phase Materials, Department of Physics, Pohang University of Science and Technology, Pohang 37673, Korea*

[5] *Rutgers Center for Emergent Materials and Department of Physics and Astronomy, Rutgers University, Piscataway, New Jersey 08854, USA*

**Abstract**

Perfect circular dichroism has been observed in the Raman scattering by the optical phonons in single chiral domain NiCo$_2$TeO$_6$ crystals. The selection rules for the optical phonons are determined by the combination of the chiral structure ***C*** and the electric polarization ***P*** along the *c*-axis. These two symmetry operations are equivalent to the ferroaxial order (***C·P***)=***A***, so the observed optical phonons are referred to as "ferroaxial". For a given Raman scattering geometry the observed effect may also be described as a complete non-reciprocal propagation of the optical phonons, whose preferable $\bar{q}$-vector direction is determined by the sign of ***C·P***. The combination of Raman scattering and polarization plane rotation of the transmitted white light allows for identification of the direction of electric polarization ***P*** in mono domain chiral crystals.

*Corresponding author *vladimir.martinez@njit.edu*

Optics reveals symmetry-based properties of materials such as nonreciprocal light propagation, rotation of the polarization plane, and circular dichroism. These effects are related to the electronic band structure of materials, and the corresponding uneven contribution of the electronic bands to the anisotropy of the dielectric function tensor, or the refractive indices. One example is the polarization plane rotation of light due to the chiral structure of sugar. This effect does not require translation symmetry and thus, polarization plane rotation can be observed in solutions containing chiral molecules, or in any system with a broken mirror symmetry. Chirality is also intrinsic to numerous single crystals, including quartz, tellurium and tellurium-based compounds. Many of these systems demonstrate intriguing topological properties, where certain electronic and spin excitations are protected by the chiral symmetry.

The recently discovered ferroaxial order in crystals [1,2,3] has many similarities with chirality: it is also characterized by the handedness of ferroaxial domains ***A***. Ferroaxial order is potentially important for spintronic devices and, in particular, for supporting longitudinal spin-currents that are protected by the handedness [4]. Similar to chiral systems, ferroaxial crystals, such as $NiTiO_3$ and $MnTiO_3$, rotate the polarization plane of the transmitted light. However, there is a fundamental difference: chirality, which is a pseudo-scalar, does not change upon a z→−z transformation, *i.e.*, upon flipping the crystal upside down, while the ferroaxial domains change sign from +*A* to –*A* under the z→−z transformation, so the ferroaxial order *A* is equivalent to an axial vector [1].

The lattice dynamics, or behavior of the optical phonons in chiral and ferroaxial crystals have attracted a significant interest recently. Two complementary concepts are under discussion: phonon behavior in the chiral/ferroaxial systems and the phonon's influence on the chiral/ferroaxial orders [5,6,7]. Work on phonons in chiral crystals, which are available as mono-domains, has continued for many decades [8,9]. Known ferroaxial crystals usually consist of micron-size domains [1], which complicate optical studies. Still, micro-Raman spectroscopy is promising for the studies of chiral and ferroaxial orders and their behavior at the level of ionic displacements in the unit cell. In several recent Raman scattering studies of chiral crystals [10,11], the so-called "chiral phonons" reveal tiny frequency shifts combined with the constant intensity of the phonon peaks when measured using circularly polarized light. In our studies of tellurium, we were not able to reproduce such phonon frequency shifts [12]. Nevertheless, we decided to study another chiral system of $NiCo_2TeO_6$. One of its peculiar properties is the electric polarization along the *c*-axis and availability of mm-size mono-domain crystals with both signs of chirality. From the start we relied on predictions of the perturbation theory, where the first-order effects caused by crystal chirality, are expected to be observed as changes of the selection rules, or phonon intensity. The phonon frequency shift we expected to be weak since those are allowed only in the second order of the perturbation theory. This has been recently reported for the chiral phonons in quartz [9].

At temperatures below $T_C$>1000 K, single crystals of $NiCo_2TeO_6$ have a defined electric polarization along the *c*-axis, +*P* and –*P*, where polarization is pinned to the crystal during the crystal growth. As shown with neutron powder diffraction, $NiCo_2TeO_6$ is isostructural to the parent $Ni_3TeO_6$, both of which have a mono-polar and mono-chiral *R*3 structure with 10 atoms in a rhombohedral unit cell [13,14]. $NiCo_2TeO_6$ has a magnetic transition near $T_N$ ~ 50 K. In contrast to $Ni_3TeO_6$, where spins of the magnetic ions $Ni^{2+}$ order collinearly along the *c*-axis, the magnetic structure of $NiCo_2TeO_6$ is incommensurate and helical with spins perpendicular to the axis of the helix. The helical magnetic structure consists of spins aligned in the *ab*-plane which stack helically along the *c*-axis with an incommensurate propagation vector $k$=(0,0,1.5±0.2973). Optical phonons and magnons have been previously studied in $NiCo_2TeO_6$ using Raman scattering and IR spectroscopy at a temperature range from 5 K to 300 K but only with linearly polarized light [15]. No effects due to structural or spin chirality have been reported.

For our Raman experiments, several mono-domain single crystals of $NiCo_2TeO_6$ with a clearly defined left or right-hand chirality have been grown, where +*C* and –*C* are two pseudo-scalars which describe the handedness of chirality. We selected ten $NiCo_2TeO_6$ samples with the lateral size of 1×1 mm$^2$ and thickness of 0.1 mm. The sign of chiral domains has been confirmed by the polarization plane rotation of the transmitted visible light measured from both sides of the flake-like crystals with the main sample plane being perpendicular to the $c$-axis. One twin crystal with a clear boundary between two chiral domains has been also identified for our experiments.

The micro-Raman experimental setup was equipped with an optical LHe-flow Oxford Microstat, a 532 nm solid state laser, a CCD detector, a 1800 g/mm grating, and a set of circular and linear polarizers. Samples were mounted in vacuum on a cold finger. A narrow-band 532 nm circular polarizer (retarder) was used for excitation, while a broadband achromatic circular polarizer (compensator) was used for analysis. All experiments were carried out in the backscattering Raman configurations $c(\sigma\pm,\sigma\pm)\bar{c}$, where the first $c$ and the last $\bar{c}$ symbols indicate the excitation and the back-scattered light propagation along the $c$-axis. The symbols in the brackets $\sigma+$ and $\sigma-$ are two helicities, or circular polarizations, of the incoming and scattered light. Due to controversial notations for circular polarization in literature, we provide more details about polarizer notation and calibration in [12]. Here we briefly mention that $c(\sigma+,\sigma+)\bar{c}$ and $c(\sigma-,\sigma-)\bar{c}$ are the configurations where the elastic, or Rayleigh, scattering is observed, while $c(\sigma+,\sigma-)\bar{c}$ and $c(\sigma-,\sigma+)\bar{c}$ correspond to "cross-circular" configurations, where spin-flip processes and magnon peaks have been detected in various Raman experiments [16, 17]. Note that in crystals without chirality or with time-reversal, no dichroism is possible and, correspondingly, the Raman configurations are equivalent for the interchange between $\sigma+$ and $\sigma-$.

**Figure 1** presents the Raman scattering spectra for one of the selected NiCo$_2$TeO$_6$ crystals. At both, $T$=300 K and $T$=5 K, a strong circular dichroism was observed between cross-circular polarizations, while the two spectra measured in the two parallel configurations are almost identical. In particular, the strongest circular dichroism was observed for the $E$–symmetry optical phonon at 325 cm$^{-1}$, which appears in $c(\sigma-,\sigma+)\bar{c}$, while this phonon is not visible in the orthogonal, $c(\sigma+,\sigma-)\bar{c}$ configuration. Other phonons also show a certain degree of the circular dichroism, which is summarized in **Table I**.

The very unusual phonon dichroism was studied in details using rotating retarder and rotating compensator measurements. In rotating retarder experiments, the polarization of the laser varies, while in the rotating compensator experiment, the polarization of the scattered light is varied. The phonon intensity as functions of the circular polarizer angle $\theta$ are shown in **FIG. 2**. When retarder rotates [**FIG.2(a)**], the laser polarization changes from circular to elliptical, and then back to linear four times, which is indicated by the symbol $\theta$ in this notation $c(\theta,\sigma+)\bar{c}$. The 360° rotation of the retarder corresponds to two complete rotations of the polarization state on the Poincare sphere, so the experimental data in **FIG. 2** must have a 180° invariance. The experimental points in **FIG.2** were fitted using the empirical $I(\theta) = I_0 \cos^2(\theta)$ function. The phonon at 325 cm$^{-1}$ is strong in $c(\theta,\sigma+)\bar{c}$ and is practically invisible for any value of $\theta$ in $c(\theta,\sigma-)\bar{c}$. We also show the simultaneously measured data for the neighboring phonon at 380 cm$^{-1}$ [**FIG. 2(a)**]. The two nearly symmetric "figure-8" dependences are expected for conventional phonons in non-chiral materials. **FIG.2(b)** shows the rotating compensator data for two phonons in $c(\sigma-,\theta)\bar{c}$ and $c(\sigma+,\theta)\bar{c}$. Again, the phonon at 380 cm$^{-1}$ shows the conventional dependencies, while we still have the single figure-8 dependence for the phonon at 325 cm$^{-1}$ with the maxima of intensity corresponding to $c(\sigma-,\sigma+)\bar{c}$ ($\theta = 135°$ and 315°) and the minima in $c(\sigma-,\sigma-)\bar{c}$ ($\theta = 45°$ and 225°). The combination of rotating retarder and rotating compensator data constitute a complete set of Raman polarimetry. This rigorous approach demonstrates that the phonon at 325 cm$^{-1}$ is indeed forbidden in one of the circular configurations and it is not rotated away into another polarization configuration.

The strongly dichroic dependencies of the phonon at 325 cm$^{-1}$ and other modes listed in **Table I** were measured at both, $T=5$ K and $T=300$ K. No significant changes were observed in the phonon spectra when the sample was cooled below the antiferromagnetic transition at $T_N=50$ K. Thus, the observed phonon dichroism is due to the crystal structure of NiCo$_2$TeO$_6$. Before attributing this effect solely to chirality, an important verification step is necessary, namely the comparison between the left- and right-chiral crystals with opposite +$C$ and −$C$ domains. However, the first round of measurements for our collection of several crystals with the same $C$-domain handedness showed a random switching of the Raman selection rules for the 325 cm$^{-1}$ phonon upon flipping the crystal's faces with respect to the laser beam. To explain the phonon dichroism, we should consider the direction of electric polarization ***P*** in each measured sample, which are not only single-chiral, but also single-***P*** domain crystals. While the sample flip does not change chirality, it does change the sign of ***P*** with respect to the direction of light propagation, e.g., with respect to the $\vec{k}$-vector of the laser excitation. At the same time, ***P*** changes direction with respect to the scattered phonon's $\vec{q}$-vector upon flipping the crystal. We can see that the phonon scattering selection rules cannot be attributed to chirality only, but to the combination of chirality and polarization, or to the sign of $C·P$. The latter has the same symmetry as the ferroaxial order ***A***, which is known to have an in-plane handedness ($\vec{A}$ is an axial vector) and changes sign upon flipping the sample face [18]. Thus, the observed phonons in NiCo$_2$TeO$_6$ are not "chiral" but they should be referred to as "ferroaxial" based on the crystal symmetry and their interaction with the circularly polarized light. The main motif of the dynamic ion displacements corresponding to the ferroaxial phonon at 325 cm$^{-1}$ is shown in **FIG. 3**.

The ferroaxial nature of the optical phonons has been additionally confirmed using the as-grown twin sample with two adjusted +$C$ and –$C$ chiral domains, where our measurements did not show any changes of the Raman selection rules on both sides with respect to the clearly visible chiral domain boundary. In the two chiral domains with the opposite +$C$ and –$C$ chirality, we had the opposite direction of polarization: (+$C·$ –***P***) on one side of the domain wall and (–$C·$ +***P***) on the other side, both producing the same sign of ***A***, and, hence, the same selection rules for the ferroaxial optical phonons. A combination of (a) and (c) in FIG. 3 corresponds to the twin sample. Also note, that the phonon at 325 cm$^{-1}$ participates in the Raman scattering only if its $\vec{q}$-vector coincides with one, let us say −***P***, direction of electric polarization. This effect may be also described as a ($C·P$) induced non-reciprocal phonon propagation in the Raman process. A similar theoretical prediction of a "phonon diode" was made recently for chiral materials [19] but the effects related to the crystal electric polarization were not included into consideration.

Dichroism of the ferroaxial phonons can be used to identify the direction of electric polarization in small crystals. In a combination with the rotation of the linear polarization plane, the Raman technique can help to select identical single-domain polar-chiral crystals to form large mosaic samples for inelastic neutron scattering or other techniques requiring large quantities of identical samples. Such a strong difference between $c(\sigma-,\sigma+)\bar{c}$ and $c(\sigma+,\sigma-)\bar{c}$ configurations has been previously observed in spin-flip Raman experiments in semiconductors in a strong magnetic field $B$, where the non-equivalence between two spin-flip channels is determined by the direction of $B$ with respect to the laser light propagation [17]. Thus, we can also conclude that the combination of $C$ and $P$ in materials is like the presence of an axial vector $\vec{B}$ in the Raman scattering experiments.

Our study of the dynamic dichroism in $NiCo_2TeO_6$ crystals would be incomplete if we did not describe the magnon spectra measured with the circularly polarized light. It has been shown [15] that below $T_N$=50 K, a magnon peak appears at 18 cm$^{-1}$ in crossed-linear configuration $c(x,y)\bar{c}$, with $x\|a$ and $y\perp a$. As expected, the crossed circular $c(\sigma-,\sigma+)\bar{c}$ and $c(\sigma+,\sigma-)\bar{c}$ configurations also show the same magnon peaks (see **FIG. 4)** at the temperature below $T_N$=50 K. Interestingly, a slight difference of about 7% was detected between $c(\sigma-,\sigma+)\bar{c}$ and $c(\sigma+,\sigma-)\bar{c}$ for the magnon at 18 cm$^{-1}$. The rotating retarder data show two slightly different "figure-8" dependencies for $c(\theta,\sigma+)\bar{c}$ and $c(\theta,\sigma+)\bar{c}$. So, the magnon selection rules in the Raman process are also affected by the electric polarization $P$ and chirality. One can envision it as two similar spin-waves propagating along the $c$-axis with nearly the same energy and each of them is excited by the corresponding circular polarization of the laser light with slightly different probability. In addition, we found a broad magnon-like spectrum around 50 cm$^{-1}$ and a sharp peak at 87 cm$^{-1}$, both measured preferably in the parallel $c(\sigma+,\sigma+)\bar{c}$ and $c(\sigma-,\sigma-)\bar{c}$ configurations. We attribute those modes at 50 cm$^{-1}$ and 87 cm$^{-1}$ to electromagnons whose oscillator strength is probably influenced by an interaction with the neighboring optical phonon at 160 cm$^{-1}$. The latter is a conventional phonon being strong in the parallel circular and linear polarizations. The electromagnon at 87 cm$^{-1}$ shows 12% dichroism in a favor of the $c(\sigma-,\sigma-)\bar{c}$ configuration over $c(\sigma+,\sigma+)\bar{c}$.

In conclusion, we introduce a new class of ferroaxial optical phonons in the polar and chiral $NiCo_2TeO_6$ crystals. The phonon selection rules in Raman scattering are determined by the simultaneous combination of the ($C\cdot P$) symmetries, which is equivalent to the ferroaxial order $A$. The 325cm$^{-1}$ phonon's $q$-vector depends on the direction of electric polarization along the $c$-axis. This effect may be described as a $C\cdot P$ induced non-reciprocal phonon propagation in the Raman process. The combination of the Raman scattering, and polarization rotation of the transmitted light allows determination of the direction of electric polarization $P$. As much as phonons in chiral materials interact with spin systems and are controlled by magnetic fields; ferroaxial phonons can provide an extra degree of materials control through the electric polarization, thereby increasing the magnetoelectric coupling and improving the understanding of chiral and ferroaxial phase formation in a broad class of magnetic and topological materials

**Acknowledgements -** The NSF MPS-ASCEND Award #2316535 supported the Raman scattering experiments and data analysis by V.A.M. The Raman scattering and sample growth by A.A.S., C.J.W., and S-W.C. were supported by the US Department of Energy DOE DE-FG02-07ER46382. The work at Postech was supported by the National Research Foundation of Korea (NRF), funded by the Ministry of Science and ICT (No. 2022M3H4A1A04074153). The sample growth at NJIT by Y.G. and J. Y. was supported by DOE DE-SC0021188. F.L. acknowledges support from the Swiss National Science Foundation (SNSF) through an Early Postdoc Mobility Fellowship #P2FRP2-199598. Work at the National Synchrotron Light Source II at Brookhaven National Laboratory was funded by the DOE DE-AC9806CH10886. Use of the 22-IR-1(FIS) beamline was supported by the NSF EAR−2223273 (Synchrotron Earth and Environmental Science, SEES) and Chicago/DOE Alliance Center (CDAC) under the DOE-NNS cooperative agreement DE-NA-0003975. The authors are thankful to G.L. Carr and C.C. Homes at NSLS-II for useful discussions.

**FIGURES**

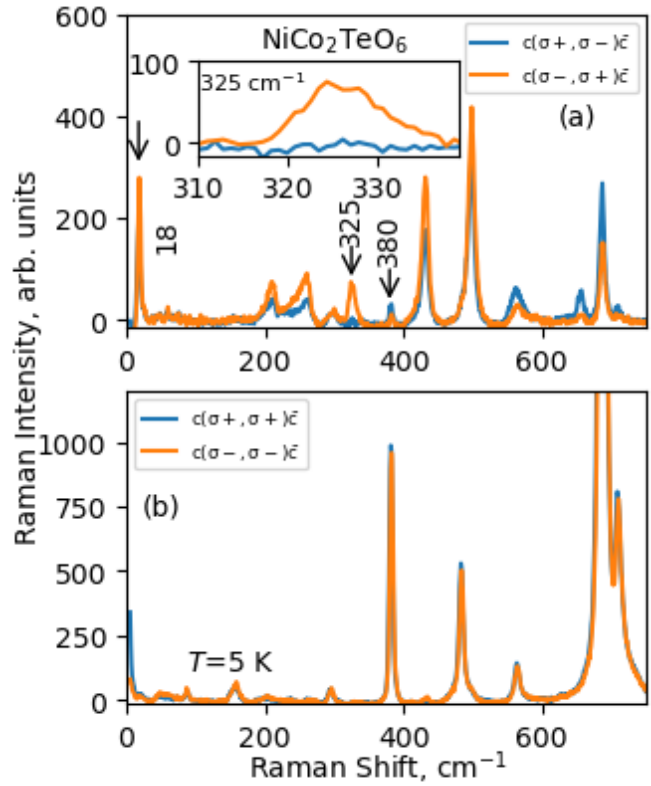

**FIG. 1** (a,b) Raman spectra in a NiCo$_2$TeO$_6$ mono-chiral domain crystal measured in (a) two crossed circular configuration: $c(\sigma-,\sigma+)\bar{c}$ and $c(\sigma+,\sigma-)\bar{c}$. Inset demonstrates the complete circular dichroism for the ferroaxial phonon at 325 cm$^{-1}$. Other phonons demonstrate partial dichroism. (b) Two parallel circular configurations: $c(\sigma-,\sigma-)\bar{c}$ and $c(\sigma+,\sigma+)\bar{c}$, where all phonons have nearly the same intensity.

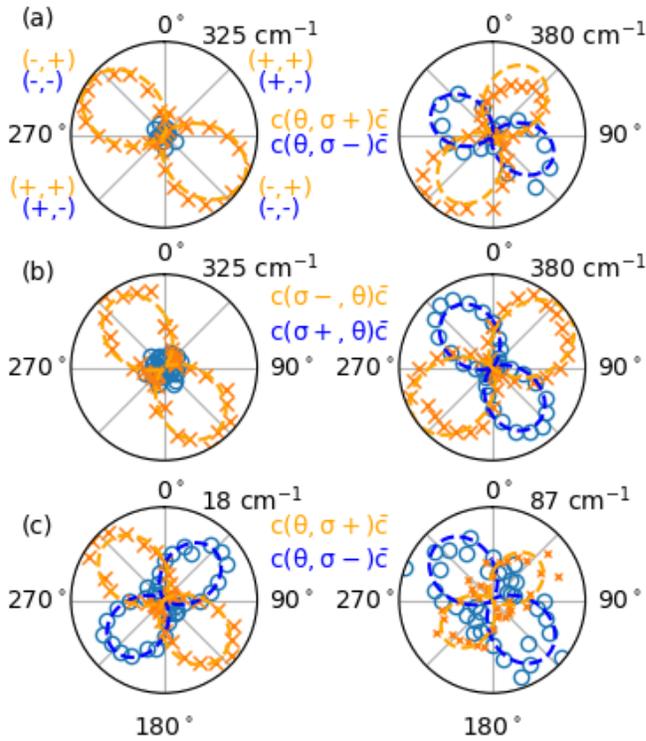

**FIG. 2** (a,b) Angular dependence of intensities for two phonons, the ferroaxial one at 325 cm$^{-1}$ on the left and the conventional phonon at 380 cm$^{-1}$ on the right measured using the rotating retarder $c(\theta,\sigma\pm)\bar{c}$ in (a) and rotating compensator $c(\sigma\pm,\theta)\bar{c}$ in (b). Experimental data are compared to the empirical functions $I(\theta) = I_0 \cos^2(\theta)$, which are shown with dashed curves. The angle $\theta$ corresponds to rotation of the circular polarizers by 360° with the step 10°. (c) Angular dependencies for the magnon at 18 cm$^{-1}$ on the left and electromagnon at 87 cm$^{-1}$ on the right using the rotating retarder $c(\theta,\sigma\pm)\bar{c}$.

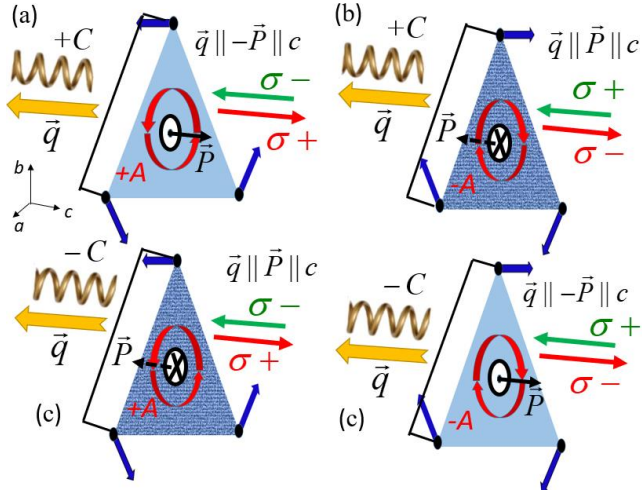

**FIG. 3** Schematics of the Raman scattering experiments with directions of the laser beam (green arrow) and the backscattered Raman signal (red arrow) and the two crystal faces, where the ferroaxial phonon at 325 cm$^{-1}$ is strong. The phonon $q$-vector is shown with yellow arrows. The corresponding ferroaxial dynamic distortions in the $a$-$b$ plane are schematically shown with the bold bleu arrows. The electric polarization vector $\vec{P}$ is shown along the $c$-axis with black arrows. The panels (a) and (b) correspond to two opposite $a$-$b$ faces of the same mono-domain sample with respect to the laser beam. Upon turning the sample upside down, the chirality $+C$ remains unchanged, while polarization $\vec{P}$ is flipped. As a result, the ferroaxial dynamic distortion $A$ also changes the sign. (c) and (d) are the two measured complementary geometries with negative $C$. In the twin sample a combination of (a) and (c) has been observed.

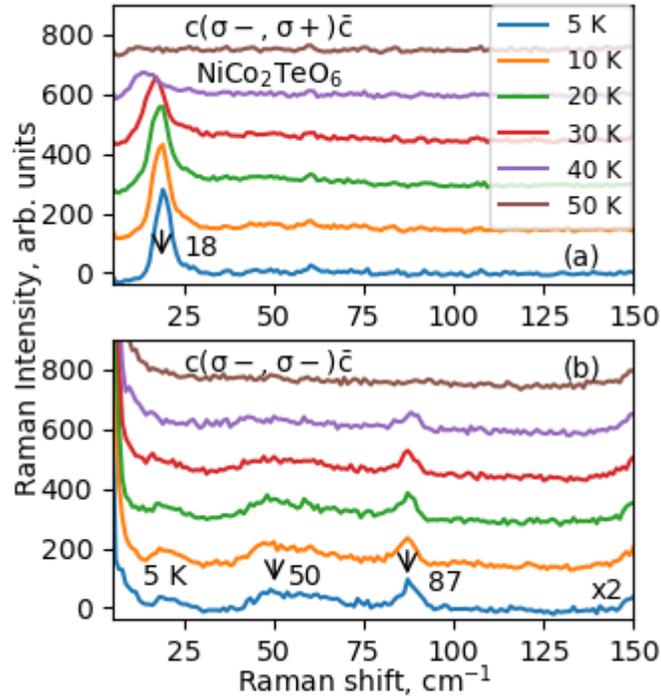

**FIG. 4** Low-frequency Raman spectra in one of the $NiCo_2TeO_6$ mono-chiral domain crystals measured at several temperatures for $T<T_N=50$ K. (a) The crossed circular configuration, spectra in $c(\sigma-,\sigma+)\bar{c}$ are dominated by the magnon at 18 cm$^{-1}$. (b) Peaks of electromagnons in the parallel circular configuration at 50 cm$^{-1}$ and 87 cm$^{-1}$.

**TABLE I.** Circular polarization of the Raman scattering intensity for several magnon and phonon peaks measured in the crossed circular configurations as $\rho = [I_0(-,+) - I_0(+,-)] / [I_0(-,+) + I_0(+,-)]$, where the signs of $\sigma\pm$ are shown in brackets and $I_0$ is the intensity for the corresponding configuration. A weak phonon at 657 cm$^{-1}$ is marked with "w".

|  | Magnon | | Phonons | | | | |
|---|---|---|---|---|---|---|---|
| Frequency, cm$^{-1}$ | 18 | 210 | 258 | 325 | 431 | 498 | 657w |
| $\rho$ | +0.07 | +0.25 | +0.22 | +1 | +0.16 | +0.11 | −0.64 |